\documentclass[11pt,letterpaper]{article}
\usepackage[letterpaper]{geometry}
\usepackage{acl2016}
\usepackage{times}
\usepackage{latexsym}
\usepackage{amsmath}
\usepackage{multirow}
\usepackage{url}
\usepackage{amssymb}
\usepackage{adjustbox}
\usepackage{cite}

\makeatletter
\newcommand{\@BIBLABEL}{\@emptybiblabel}
\newcommand{\@emptybiblabel}[1]{}
\makeatother
\usepackage[hidelinks]{hyperref}
\usepackage{bm} 

\usepackage[english]{babel}

\usepackage{graphicx} 
\graphicspath{{figures/}} 
\usepackage{xcolor} 
\usepackage{caption}
\usepackage{subfigure}

\usepackage{booktabs}

\aclfinalcopy 

\title{HLA class I binding prediction via convolutional neural networks}

\author{Yeeleng S. Vang and Xiaohui Xie \\
Department of Computer Science\\
University of California, Irvine, CA 92697, USA\\
{\tt \{ysvang,xhx\}@ics.uci.edu}}

\date{}

\begin{document}
\maketitle
\begin{abstract}
Many biological processes are governed by protein-ligand interactions.  One such example is the recognition of self and nonself cells by the immune system. This immune response process is regulated by the major histocompatibility complex (MHC) protein which is encoded by the human leukocyte antigen (HLA) complex. Understanding the binding potential between MHC and peptides can lead to the design of more potent, peptide-based vaccines and immunotherapies for infectious autoimmune diseases.

We apply machine learning techniques from the natural language processing (NLP) domain to address the task of MHC-peptide binding prediction.  More specifically, we introduce a new distributed representation of amino acids, name HLA-Vec, that can be used for a variety of downstream proteomic machine learning tasks. We then propose a deep convolutional neural network architecture, name HLA-CNN, for the task of HLA class I-peptide binding prediction. Experimental results show combining the new distributed representation with our HLA-CNN architecture acheives state-of-the-art results in the majority of the latest two Immune Epitope Database (IEDB) weekly automated benchmark datasets. We further apply our model to predict binding on the human genome and identify 15 genes with potential for self binding. Codes are available at \href{https://github.com/uci-cbcl/HLA-bind}{https://github.com/uci-cbcl/HLA-bind}.
\end{abstract}

\section{Introduction}
The major histocompatibility complex (MHC) are cell surface proteins used to bind intracellular peptide fragments and display them on cell surface for recognition by T-cells \cite{Janeway01}.  In humans, the human leukocyte antigens (HLA) gene complex encodes these MHC proteins.  HLAs displays a high degree of polymorphism, a variability maintained through the need to successfully process a wide range of foreign peptides \cite{Jin03, Williams01}.  

The HLA gene lies in chromosome 6p21 and is comprised of 7.6Mb \cite{Simmonds07}. There are different classes of HLAs including class I, II, and III corresponding to their location in the encoding region. HLA class I is one of two, the other being class II, primary classes of HLA.  Its function is to present peptides from inside cells to be recognized either as self or nonself as part of the immune system. Foreign antigens presented by class I HLAs attracts killer T-cells and provoke an immune response. Similarly, class II HLAs are only found on antigen-presenting cells, such as mononuclear phagocytes and B cells,  and presents antigen from extracellular proteins \cite{Ulvestad94}. Unlike class I and II, class III HLAs encode proteins important for inflammation. 

The focus of this paper is on HLA class I proteins. As these molecules are highly specific, they are able to bind with only a tiny fraction of the peptides available through the antigen presenting pathway \cite{Nielsen16, Yewdell99}. This specificity makes binding to the HLA protein the most critical step in antigen presentation. Due to the importance of binding, accurate prediction models can shed understanding to adverse drug reactions and autoimmune diseases \cite{Gebe02, Illing12}, and lead to the design of more effective protein therapy and vaccines \cite{Chirino04, van06}.

Given the importance of MHC to the immune response, many algorithms have been developed for the task of MHC-peptide binding prediction. The following list is by no means exhaustive but a small sample of previously proposed models. Wang et al. proposed using quantitative structure-activity relationship (QSAR) modeling from various amino acid descriptors with linear regression models \cite{Wang15}. Kim et al. derived an amino acid similarity matrix \cite{Kim09}. Luo et al. proposed both a colored and non-colored bipartite networks \cite{Luo16}. Shallow and high-order artificial neural networks were proposed from various labs \cite{Hoof09, Koch13, Kuksa15, Nielsen03}. Of these approaches, NetMHC/NetMHCpan have been shown to achieve state-of-the-art for MHC-peptide binding prediction \cite{Nielsen16, Trolle15}.

In this article, we apply machine learning techniques from the natural language processing (NLP) domain to tackle the task of MHC-peptide binding prediction. Specifically, we introduce a new distributed representation of amino acids, named HLA-Vec, that maps amino acids to a 15-dimensional vector space.  We combine this vector space representation with a deep convolutional neural network (CNN) architecture, named HLA-CNN, for the task of HLA class I-peptide binding prediction. Finally, we provide evidence that shows HLA-CNN achieves state-of-the-art results for the majority of different allele subtypes from the IEDB weekly automated benchmark datasets.

\section{Methods}

\subsection{Dataset}

\begin{figure}[t]
\centerline{\includegraphics[width=0.5\textwidth]{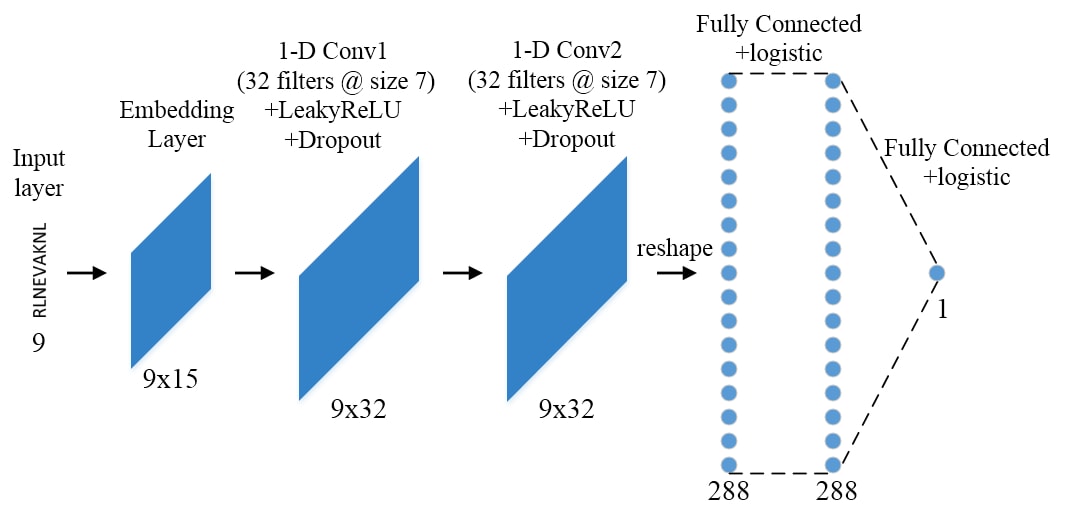}}
\caption{We illustrate our CNN architecture for MHC-peptide binding prediction of size 9-mers. The input is the peptide. The embedding layer substitues the individual amino acids with their 15-dimensional vector space representation. This is followed by two 1-dimensional convolutional layers preserving input length using 32 filters of size 7. The output of the 2nd convolutional layer is reshape into a 1-dimensional vector and is  fully connected to the next layer of the same size. This fully connected layer is then fully connected to a logistic output unit. The architecture is generalizable to allele subtypes of any length.}\label{fig:01}
\end{figure}

To control for data pre-processing variabilities, we decided to use an existing post-processed training dataset so prediction algorithms could be more directly compared. The dataset used was filtered, processed, and prepared by Luo et al. \cite{Luo16}. This dataset contained HLA class I binding data curated from four widely used, publicly available MHC datasets: IEDB \cite{Vita15}, AntiJen \cite{Toseland05}, MHCBN \cite{Lata09}, and SYFPEITHI \cite{Rammensee99}. Target indicator indicating binding or nonbinding was readily given as one of the column in the processed dataset. Peptides that contained unknown or indiscernible amino acids, denoted as "X" or "B", were removed from the dataset prior to training. Dataset was split into 70\% training set and 30\% validation set.

The test datasets were obtained from IEDB automatic server benchmark page (\href{http://tools.iedb.org/auto\_bench/mhci/weekly/} {http://tools.iedb.org/auto\_bench/mhci/weekly/}). Allele subtypes with less than 500 training examples were excluded from testing. The lack of training data is a well-known weakness of deep neural networks as the model may not converge to a solution or worst yet, may overfit to the small training set. Indicators of binding were given as either binary values or ic50 (half maximal inhibitory concentration) measurements.  Binary indicators were used directly while values given in ic50 measurements were denoted as binding if ic50 $<$ 500 nM.

\subsection{Distributed Representation}

\begin{table*}[t]
\caption{HLA-Vec, an amino acids distributed representation.\label{Tab:01}}
\begin{adjustbox}{width=1\textwidth}
{\begin{tabular}{cccccccccccccccc}\hline
Amino Acid & dim-1 & dim-2 & dim-3 & dim-4 & dim-5 & dim-6 & dim-7 & dim-8 & dim-9 & dim-10 & dim-11 & dim-12 & dim-13 & dim-14 & dim-15 \\ \hline
A & -0.1428 & 0.1384 & 0.0579 & -0.2463 & 0.3611 & 0.2930 & -0.2692 & -0.1532 & 0.0249 & -0.0250 & 0.4739 & -0.1261 & -0.1988 & 0.0109 & -0.0323\\
C & 0.1210 & -0.4359 & -0.2869 & -0.3863 & 0.2308 & 0.0282 & 0.0265 & -0.0629 & -0.0820 & 0.2711 & 0.2992 & -0.3216 & 0.0350 & -0.1972 & 0.0464\\
E & -0.0669 & 0.2425 & 0.0402 & -0.2557 & -0.0065 & 0.1357 & 0.0091 & -0.2227 & -0.2217 & 0.0975 & 0.4011 & -0.0615 & -0.0142 & -0.2462 & -0.1409\\
D & -0.0493 & 0.0821 & -0.0815 & -0.3480 & 0.0629 & 0.2017 & 0.2364 & -0.1319 & -0.0762 & 0.2678 & 0.1717 & -0.1500 & 0.0241 & -0.1535 & -0.2351\\
G & -0.1611 & 0.0342 & -0.1203 & -0.1648 & 0.2604 & 0.0739 & 0.1568 & -0.2627 & 0.0374 & 0.0746 & 0.2992 & -0.3769 & -0.1973 & -0.0003 & -0.1831\\
F & -0.1002 & -0.0711 & -0.2272 & -0.1740 & 0.2519 & 0.1076 & 0.1501 & -0.1994 & -0.0486 & 0.0774 & 0.1696 & -0.0822 & 0.2303 & -0.1641 & -0.2655\\
I & -0.0833 & 0.1719 & -0.2545 & -0.2451 & 0.1372 & 0.3516 & 0.0905 & -0.0902 & -0.1880 & 0.0864 & 0.0773 & 0.0309 & 0.1380 & -0.2591 & -0.1420\\
H & -0.1433 & -0.0003 & -0.0744 & -0.1195 & 0.3056 & 0.1037 & 0.0642 & -0.0514 & -0.1960 & 0.2619 & 0.1837 & -0.2322 & 0.1123 & -0.0715 & -0.2034\\
K & 0.0276 & 0.1958 & -0.2127 & -0.1873 & 0.0951 & 0.0930 & 0.0423 & -0.2705 & -0.1871 & 0.2797 & 0.2675 & 0.0294 & 0.0429 & -0.2417 & -0.0357\\
M & -0.1812 & 0.1389 & -0.1602 & -0.1305 & 0.3983 & 0.2286 & -0.2404 & -0.1206 & 0.1616 & 0.3216 & 0.0079 & -0.2107 & -0.0639 & -0.2552 & -0.0892\\
L & -0.0360 & 0.0927 & -0.1824 & -0.1546 & 0.2046 & -0.0066 & -0.0146 & -0.2175 & -0.1835 & 0.2056 & 0.4694 & 0.0770 & 0.0804 & -0.1424 & 0.0520\\
N & -0.1220 & 0.1053 & -0.1168 & -0.4092 & -0.0007 & 0.2341 & 0.0629 & -0.0610 & -0.1055 & 0.4091 & 0.1039 & 0.0990 & 0.2457 & -0.1396 & -0.2956\\
Q & -0.5482 & 0.0352 & 0.1479 & -0.0171 & -0.1638 & 0.0976 & -0.0539 & -0.3098 & -0.1891 & 0.0823 & 0.3988 & -0.1479 & 0.0769 & -0.2872 & -0.1089\\
P & -0.3671 & -0.1098 & -0.0392 & 0.0031 & 0.2176 & 0.3222 & 0.1557 & -0.1623 & 0.0569 & 0.1854 & 0.2744 & -0.2758 & 0.2775 & 0.0526 & 0.0957\\
S & -0.0774 & -0.0416 & -0.2532 & -0.1159 & 0.2320 & 0.0761 & -0.0995 & -0.2774 & -0.0892 & 0.2454 & 0.1238 & -0.1930 & 0.0999 & -0.1710 & -0.1671\\
R & 0.3854 & 0.1272 & -0.3518 & -0.1442 & 0.2487 & 0.0564 & 0.1701 & -0.1434 & 0.1015 & -0.0507 & 0.2773 & -0.0669 & 0.2507 & -0.0338 & -0.0685\\
T & -0.0935 & 0.0087 & -0.1558 & -0.1983 & 0.2365 & 0.2426 & 0.0244 & -0.0749 & -0.1608 & 0.0807 & 0.2357 & -0.1303 & 0.1860 & -0.1256 & -0.0830\\
W & -0.4829 & -0.0159 & 0.0106 & 0.0676 & 0.3279 & -0.1073 & -0.0050 & -0.1282 & -0.1045 & -0.0425 & 0.1982 & -0.2086 & -0.0252 & -0.4396 & -0.3651\\
V & -0.11540 & .0944 & -0.1744 & -0.0475 & 0.2863 & 0.3909 & 0.1128 & -0.1018 & -0.1815 & 0.0061 & 0.1972 & -0.1604 & 0.0812 & -0.2151 & 0.1363\\
Y & -0.1308 & -0.0410 & -0.1395 & 0.0534 & 0.3133 & 0.2197 & 0.1469 & -0.1309 & -0.3230 & 0.2696 & 0.0919 & -0.0462 & 0.0193 & -0.2942 & -0.0820\\
\hline
\end{tabular}}
\end{adjustbox}
{This table lists the 15-dimensional vector space distributed representation of amino acid trained unsupervised on HLA class I peptides of all allele subtypes and lengths from the training dataset. The dimensions are arbitrary and have no physicochemical interpretation.}
\end{table*}

Distributed representation has been successfully used in NLP to train word embeddings, the mapping of words to real-value vector space representations. More generally, distributed representation is a means to represent an item by its relationship to other items. In word embeddings, this means semantically similar words are mapped near each other in the distributed representation vector space. The resulting distributed representation can then be used much like how BLOSUM is used for sequence alignment of proteins \cite{Henikoff92} or peptide binding prediction by NetMHCpan \cite{Andreatta15}. That is, we encode amino acids with their vector space distributed representation to be useable by downstream machine learning algorithms. Other amino acid encoding includes Atchley factors \cite{Atchley05} and Kidera factors \cite{Kidera85}, both of which were constructed explicitly to summarize amino acid physicochemical properties. In the end-to-end machine learning approach we propose, the encoding is learned directly from the raw amino acid sequences in an unsupervised manner. The vector representation is obtained without any manual input, and as a result, the vector space has no explicit interpretations unlike Atchley or Kidera factors.

Recently, distributed representation had been explored for bioinformatics applications.  Specifically, trigram (sequence of 3 amino acids) protein distributed representation of size 100-dimensions was used to encode proteins for protein family classification and identifying disordered sequences, resulting in state-of-the-arts performance \cite{Asgari15}. The distributed representation was further shown to grouped trigram proteins with similar physicochemical property closer to each other by mapping the 100-dimensional space to 2-dimension.

Distributed representation approaches can be classified into two broad classes: prediction-based and count-based. Two of the most popular prediction-based, neural probabilistic language models commonly used to develop a distributed representation are the skip-gram model and continuous bag-of-words (CBOW) model \cite{Mikolov13b}. Both models are similar and is often thought of as inverse of one another. In the skip-gram model, the adjacent context-words are predicted based on the center (target) word. Conversely in the CBOW model, the center word is predicted based on adjacent context-words. 

A recently proposed distributed representation based on more traditional count-based method is GloVe \cite{Pennington14}. In this approach, the authors worked on   co-occurrence statistics explicitly and cast the problem as a weighted least square problem with the aim to minimize the difference between the inner product of each pair of word vectors and the logrithm of their co-occurrences. With certain assumption, the authors showed that the skip-gram model's cost function can be formulated equivalently to the GloVe model. However, it has been shown that prediction-based models are superior to count-based models \cite{Baroni14}, and under equal conditions where both models' hyperparameters were highly tuned, the skip-gram model consistently outperformed GloVe model on a number of NLP tasks \cite{Levy15}.

In this paper, the skip-gram model is used. The interested reader is encourage to consult the relevant references for further details of the CBOW or GloVe models. 

A short overview of the skip-gram model is given here for completeness. As originally formulated by Mikolov \cite{Mikolov13a}, in the skip-gram model, given a sequence of words $w_1, w_2, ..., w_T$, the objective is to maximize the average log probability:

\begin{equation}
\frac{1}{T} \sum_{t=1}^{T} \sum_{-c \leq j \leq c, j \neq 0} \text{log} \ p(w_{t+j}|w_j)\label{eq:01}\vspace*{-10pt}
\end{equation}\\

where $T$ is the total number of words (i.e. total number of amino acids) in the dataset, $c$ is the context window size (i.e. number of words to the right or left of the target word, and $p(w_{t+j}|w_t)$ is defined as:

\begin{equation}
p(w_O|w_I)=\frac{\text{exp} \ ({v_{w_O}^{'}}^\intercal  v_{w_I})}{\sum_{w=1}^W \text{exp} ({v_{w_O}^{'}}^\intercal  v_{w_I})}\label{eq:02}\vspace*{-10pt}
\end{equation}\\

Here, $v_w$ and $v_w^{'}$ are two vector space representations of the word $w$.  The subscripts $O$ and $I$ correspond to the output (context-words) word and input (target) word respectively. $W$ is the total number of unique words in the vocabulary. In typical NLP text corpus with large vocabulary, calculating the gradient of the log probability becomes impractical. An approximation to the log probability is obtained by replacing every $\text{log} \ p(w_O|w_I)$ with 

\begin{equation}
\text{log} \ \sigma({{v_{w_O}^{'}}^\intercal v_{w_I}}) + \sum_{i=1}^k \mathbb{E}_{w_i\sim P_n(w)} [\text{log} \  \sigma({v_{w_i}^{'}}^\intercal v_{w_I})]\label{eq:03}\vspace*{-10pt}
\end{equation}\\

where $\sigma(x) = 1/(1+\text{exp}(-x))$ and $k$ are negative samples. This was motivated by the idea that a good model should be able to differentiate real data from false (negative) ones.

By formulating protein data as standard sequence data like sentences in a text corpus, standard NLP algorithms can be readily applied. More concretely, individual peptides are treated as individual sentences and amino acids are treated as words. In this paper, the skip-gram model is used with a context window of size 5, 5 negative samples, and 15-dimensional vector space embedding. Various other dimensional size were explored, however, 15-dimensions gave the best results on 10-fold cross-validation of HLA-A*02:01 subtype. The entire post-processed dataset by Luo et al. \cite{Luo16} was used to learn this new distributed representation. The 15-dimensional vector space distributed representation, HLA-Vec, is summarized in Table \ref{Tab:01}. Experimental results indicate using our proposed HLA-Vec encoding showed performance gains over Asgari's representation, Atchley factors, or Kidera factors. Description of the experiements results can be found in the Supplementary Material.

\subsection{Convolutional neural network}

\begin{table*}[t]
\caption{Performance comparison of NetMHCpan, sNebula, and HLA-CNN on IEDB datasets.\label{Tab:02}} 
\begin{adjustbox}{width=1\textwidth}
{\begin{tabular}{|c|c|c|c|c|c|c|c|c|c|c|c|}\hline
Dataset & IEDB & HLA & Peptide length & Peptide count & Measurement &
\multicolumn{2}{c|}{NetMHCpan} & \multicolumn{2}{c|}{sNebula} & \multicolumn{2}{c|}{HLA-CNN} \\
\cline{7-12} 
2015-08-07 & 1029125 & B*27:05 & 9 & 21 & binary & 0.751 & \textbf{0.959} & \textbf{0.752} & \textbf{0.959} & 0.684 & 0.918\\
2015-08-07 & 1029061 & B*57:01 & 9 & 26 & ic50 & \textbf{0.612} & \textbf{0.943} & 0.169 & 0.575 & 0.443 & 0.807\\
2015-08-07 & 1028928 & A*02:01 & 9 & 13 & binary & \textbf{0.570} & \textbf{0.955} & 0.539 & 0.909 & \textbf{0.570} & \textbf{0.955}\\
2015-08-07 & 1028928 & B*07:02 & 9 & 12 & binary & \textbf{0.648} & \textbf{1.000} & 0.522 & 0.900 & \textbf{0.648} & \textbf{1.000}\\
2015-08-07 & 315174 & B*27:03 & 9 & 11 & binary & 0.657 & 0.893 & 0.179 & 0.607 & \textbf{0.837} & \textbf{1.000}\\
2015-08-07 & 1028790 & A*02:01 & 9 & 55 & ic50 & \textbf{0.615} & 0.574 & 0.505 & \textbf{0.778} & 0.580 & 0.681\\
2015-08-07 & 1028790 & A*02:01 & 10 & 35 & ic50 & 0.407 & 0.677 & \textbf{0.432} & \textbf{0.704} & 0.327 & 0.589\\
2015-08-07 & 1028790 & B*02:02 & 9 & 55 & ic50 & \textbf{0.582} & 0.713 & 0.372 & 0.680 & 0.426 & \textbf{0.804}\\
2015-08-07 & 1028790 & A*02:03 & 9 & 55 & ic50 & \textbf{0.539} & 0.696 & 0.477 & 0.629 & 0.373 & \textbf{0.746}\\
2015-08-07 & 1028790 & A*02:03 & 10 & 35 & ic50 & 0.208 & 0.750 & \textbf{0.419} & 0.697 & 0.307 & \textbf{0.837}\\
2015-08-07 & 1028790 & A*02:06 & 9 & 55 & ic50 & \textbf{0.630} & 0.770 & 0.510 & \textbf{0.848} & 0.578 & 0.819\\
2015-08-07 & 1028790 & A*02:06 & 10 & 35 & ic50 & 0.572 & 0.768 & 0.525 & 0.680 & \textbf{0.638} & \textbf{0.920}\\
2015-08-07 & 1028790 & A*68:02 & 9 & 55 & ic50 & 0.534 & 0.806 & 0.482 & 0.713 & \textbf{0.581} & \textbf{0.909}\\
2015-08-07 & 1028790 & A*68:02 & 10 & 35 & ic50 & 0.272 & 0.620 & 0.591 & 0.813 & \textbf{0.722} & \textbf{0.991}\\
\cline{1-12}
& & & & & Average: & .511 & .778 & .436 & .735 & .521 & .836\\
\hline
\end{tabular}}
\end{adjustbox}
{These benchmark datasets came from IEDB and encompasses the two most recent datsets. Allele subtypes with fewer than 500 training examples were excluded from these test datasets. SRCC stands for Spearman's rank correlation coefficient and AUC stands for area under the receiver operating characteristic curve.}
\end{table*}

Convolutional neural networks (CNN) have been studied since the late 1980s and have made a comeback in recent years along with renewed interested in artificial neural networks, and in particular of the deep architecture varieties. Much of the recent fervor has been spurned in part by both accessibility to large training datasets consisting of over millions of training examples and advances in cheap computing power needed to train these deep network architectures in a reasonable amount of time.  Although originally proposed for the task of image classification \cite{LeCun89, Krizhevsky12, Simonyan14}, CNN have been found to work well for general sequence data such as natural language sentences \cite{Kalchbrenner14, Kim14}. It is with this insight that we propose a convolutional neural network for the task of MHC-peptide binding prediction.

The CNN architecture we propose in this paper consists of both convolutional and fully connected (dense) layers. Convolutional layers preserve local spatial information \cite{Taylor10} and thus is well suited for studying peptides where spatial locations of the amino acids are critical for bonding. 

Our CNN model, dubbed HLA-CNN, can be seen in Fig. \ref{fig:01}. The input into HLA-CNN network is the character string of the peptide, a 9-mer peptide in this example. The input feeds into the embedding layer that substitutes each amino acid with their 15-dimensional vector space representation. The output encoding is a 2-dimensional matrix of size 9x15. The vector space matrix is then 1-dimensionally convolved with 32 filters of length (rows) 7 and returns the same output sequence length as input, resulting in a matrix of size 9x32. 1-dimensional convolution automatically constrains the current filter's column size to be identical to the incoming input matrix's column size. Therefore each of the 32 filters in the conv1 layer are of size 7x15, and in the conv2 layer are of size 7x32. With appropriate zero-padding of the input matrix, the same output sequence length, e.g. 9, is returned. More formally, the 1-d convolution formula is defined as:

\begin{equation}
G[i,k] = F_k \ast H = \sum_{u} \sum_{v=0}^{M} F_k[u,v]H[i-u,M-v] \label{eq:04}\vspace*{-10pt}
\end{equation}\\

where $F_k$ is the $k^{th}$ filter, $H$ is the input matrix, $G$ is the output matrix, $M$ is the column size of $H$ minus 1, and $u$ ranges from $-\lfloor \frac{filter \ length}{2} \rfloor$ to $\lfloor \frac{filter \ length}{2} \rfloor$.

The activation unit use is the leaky rectified linear units (LeakyReLU) with default learning rate of 0.3.  LeakyReLU is similar to rectified linear units except there is no zero region which results in non-zero gradient over the entire domain \cite{Maas13}. Dropout is used after each of the convolutional layers. Dropout acts as regularization to prevent overfitting by randomly dropping a percentage of the units from the CNN during training \cite{Srivastava14}. This has the effect of preventing co-adaptation between neurons, the state where two or more neurons detect the same feature. In our architecture, the dropout percentage is set to 25\%.  The output then feeds into a second convolutional layer with the same filter length, activation unit, and dropout as the first convolutional layer. The 9x32 matrix outputted by the second convolutional layer is reshaped into a single 1-D vector of size 288 which is fully connected to another layer of the same size with sigmoid activation units.  This dense layer is then fully connected to a logistic regression  output unit to make a prediction.

The loss function used is the binary cross entropy function and the optimizer used is the Adam optimizer with learning rate 0.004.  We use a variable batch size instead of a fixed one, choosing instead to force all allele subtypes to have 100 batches no matter the total number of training samples of each subtype.  The convolutional layers' filters are initialized by scaling a random Gaussian distribution by the sum of edges coming in and going out of those layers \cite{Glorot10}. Finally, the embedding layer of HLA-CNN is initialized to the previously learned HLA-Vec distributed representation with the caveat that the embedding layer is allowed to be updated during the supervised binding prediction training for each allele subtypes. This allows for the distributed representation to be fined-tuned for each allele subtypes uniquely and for the task of peptide binding specifically. The number of epoch was less important as we arbitrarily set max epoch to 100 but enforce early stoppage if the loss function stops improving for 2 consecutive epochs. Solutions were found to have converged under 40 epochs for all test sets. 

The dataset was most abundant in 9-mer HLA-A*02:01 allele (10547 samples) therefore this specific 9-mer subtype was used for network architectural design and hyperparameter tuning.  Dataset split of 70\% training and 30\% validation was used to determine the optimal architecture and hyperparamters. While the network architecture was designed using a single allele subtype of length 9, HLA-CNN framework is robust enough to accept and make prediction for allele subtypes of any length. 

Each test datasets of different allele subtypes and peptide lengths are treated as completely separate tests. For a specific test dataset, the training dataset is filtered on the allele subtype and peptide length.  The resulting smaller training subset is then used to train the HLA-CNN model. Due to the random nature of initialization in the deep learning software framework used, five prediction scores are made for each test sets. The final prediction used for evaluation purposes is taken as the average predicted score of the five predictions.  Two commonly used evaluation metric for peptide binding prediction task are the Spearman's rank correlation coefficient (SRCC) and area under the receiver operating characteristic curve (AUC). The state-of-the-art NetMHCpan \cite{Andreatta15, Trolle15}, a shallow feed forward neural network, and a more recently developed bipartite network-based algorithm, sNebula  \cite{Luo16}, will be used to compared the performance of our proposed HLA-CNN prediction model.

\begin{figure*}[bth]
\centerline{\includegraphics[width=\paperwidth]{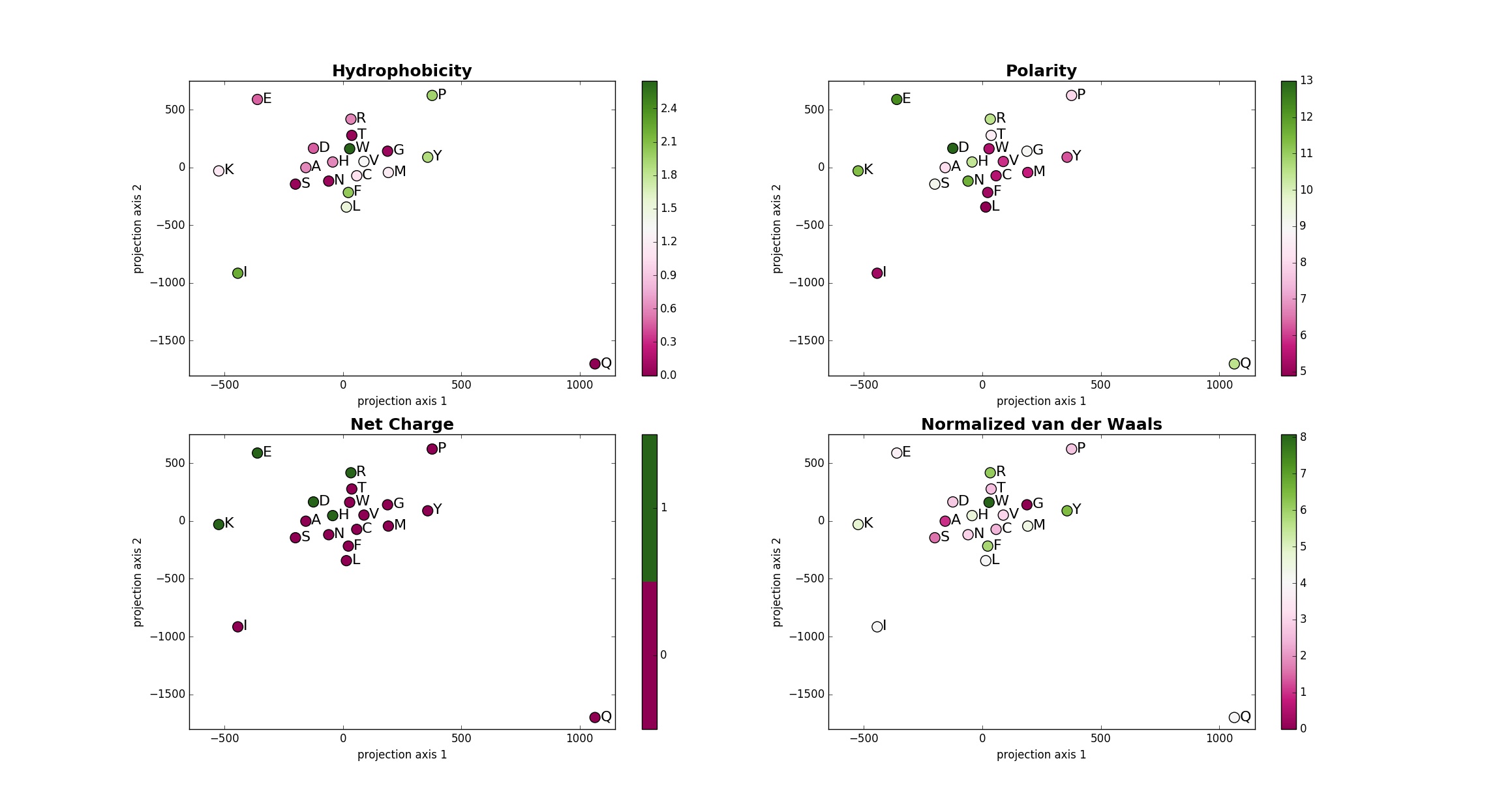}}
\caption{In these plots, each point represents the 2-D mapping of an amino acid from the 15-dimensional distributed representation using t-SNE. The color indicates the scale of each physicochemical property. Each amino acid is labeled with its one-letter code.}\label{fig:02}
\end{figure*}

\section{Results}

We have introduced the HLA class I dataset. We formulated the HLA class I peptide data as an equivalence of text data used in NLP machine learning tasks. We have proposed a model to learn a vector space distributed representation of amino acids from this HLA class I dataset. We have also described our deep learning method and how it takes advantage of this new distributed representation of amino acids to solve the problem of HLA class I-peptide binding prediction. Next, we show the result of the learned distributed representation followed by the performance of our model against the state-of-the-art prediction model and another recently developed model.

\subsection{Distributed Representation}

The 15-dimensional distributed representation of amino acids is shown in Table \ref{Tab:01}. Each of the 15 dimensions on their own have no explicit physicochemical interpretation, unlike in Atchley factors \cite{Atchley05} or Kidera factors \cite{Kidera85}. They are simply the result of the algorithm and our choice of embedding size for the representation. 

To see if the learned, 15-dimensional distributed representation of the twenty amino acids was able to capture any interesting pattern, we reduce the 15-dimensional vector space to a visualizable 2-dimensional representation using a dimension reduction technique called t-distributed stochastic neighboring embedding (t-SNE) \cite{Maaten08}. t-SNE is capable of preserving local structure of the data, e.g. points closer to each other in the original, high-dimensional space are grouped closer together in the low 2-dimensional space. We color this low dimensional representation with various physicochemical properties to see if any pattern can be discerned using this unsupervised machine learning technique.

In Fig \ref{fig:02}, we see the 2-D visualization of HLA-Vec colored by various physicochemical properties, including hydrophobicity, normalized van der waals volume, polarity, and net charge \cite{Asgari15} from the Amino acid Physicochemical properties Database (APDbase) \cite{Mathura05}. As can be seen, there are some structure in the graphs for hydrophobicity, polarity, and net charge; factors important for covalent chemical bonding.  The clusters of magenta-colored amino acids are almost separable from clusters of green-colored amino acids with the exception of a few outliers. This gives validation to distributed representation as an effective technique to automatically learn encoding that is able to preserve some important physicochemical properties without explicitly constructing such an encoding by hand.

\begin{table}[b]
\centering
\caption{Average benchmark datasets performance with model ablations. We find that the CNN is most important.\label{Tab:03}} {\begin{tabular}{ccc}\hline
 & SRCC &  AUC  \\ \hline
HLA-CNN & .521 & .836\\
-Distributed Rep. & .521 & .819\\
-CNN  & .513 & .818\\
\hline
\end{tabular}}
\end{table}

\subsection{HLA-peptide binding prediction}

The results of our HLA-CNN prediction model against NetMHCpan and sNebula on the two latest IEDB benchmarks are shown in Table \ref{Tab:02}. As AUC is a better measure of the goodness of binary predictors compared to SRCC, for evaluation purposes between models, we say one algorithm is superior to another if it scores higher on the AUC metric. 

On these latest IEDB benchmark datasets, our algorithm achieved state-of-the-art results in 10 of 15 (66.7\%) test datasets. This is in contrast to NetMHCpan, which acheived state-of-the-art results in only 4 out of 15 (26.7\%) and sNebula in 4 out of 15 (26.7\%). In the 10 allele subtypes where our model achieved state-of-the-art results, our model averaged a 9.3\% improvement over the previous state-of-the-art. 

As the binary cross-entropy loss function for this binding prediction problem operates on binary-transformed indicator values, any sort of ranking information encoded in ic50 binding measurements are loss in the objective and is a secondary task. Indeed, we observed no strong correlation or monotonicity between SRCC and AUC. Our algorithm scored highest for the SRCC metric on 7 of 15 test sets. NetMHCpan scored highest on 7 test sets as well and sNebula highest on 3 test sets. However, on average performance over all subtypes, our model gained a modest 1\% improvement over netMHCpan.

\begin{figure}[bt]
\centerline{\includegraphics[width=0.5\textwidth]{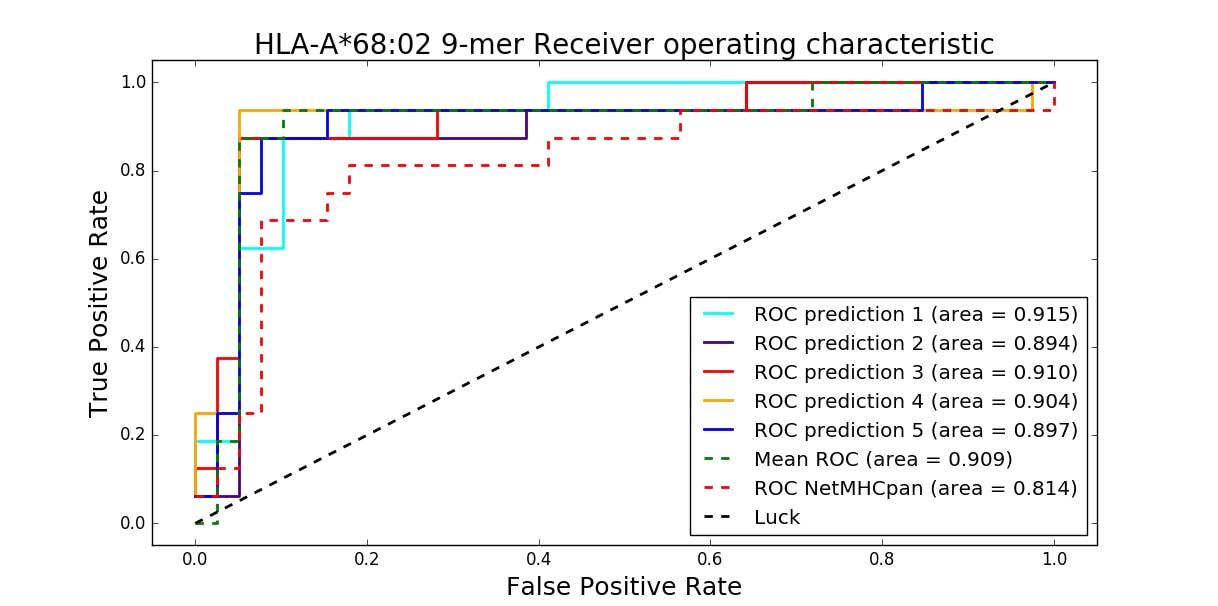}}
\caption{The ROC curves for HLA-A*68:02 9-mer test set is shown for all five predictions and their mean compared against NetMHCpan. Our model shows improvement across all five predictions and the average prediction.}\label{fig:03}
\end{figure}

In Fig. \ref{fig:03}, the ROC curves are shown for all five predictions of the HLA-A*68:02 9-mer subtype as an example of the improvement our model gives over the previous state-of-the-part. As can be seen, all five curves are outperforming NetMHCpan's curve at almost all thresholds. 

The results suggests that HLA-CNN can accurately predict HLA class I-peptide binding and outperforms the current state-of-the-art algorithms.  The results also confirmed that the hyperparamters of HLA-CNN learned on the HLA-A*02:01 9-mer subtype generalizes well to cover a variety of other allele subtypes and peptide lengths, demonstrating the robustness of our algorithm.

\subsection{Model Ablations}

In order to understand whether the distributed representation or the CNN was responsible for the performance of HLA-CNN, we perform a model ablation analysis where we remove one component of our algorithm at a time. The result shown in Table \ref{Tab:03} is of average SRCC and AUC scores over the different allele subtypes. The table indicates that the CNN is most important. In the -CNN architecture, we run a one hidden layer (66 units) neural network like NetMHCpan \cite{Nielsen16} using the HLA-Vec distributed representation encoding and allow for fine-tuning during training step. In the -Distributed Rep. architecture, we use a sparse, one-hot encoding with the CNN. Performance for each allele subtypes under these two models are available in the Supplementary Material.

\subsection{Run-time Analysis}

Though we did not do so, due to the relatively small peptide binding dataset compared to ones typically seen in NLP, both HLA-Vec and HLA-CNN can be parallelized on a GPU for faster computation. Using a single thread of a quad 3.5GHz Intel Core i7 machine, HLA-Vec learned on the entire dataset took 33 seconds.  HLA-CNN trained on the largest allele subtype, A*02:01 9-mers, took less than 10 minutes to run.

\begin{table}[t!]
\centering
\caption{Top 15 human 9-mers predicted by HLA-CNN to bind to HLA-A*02:01.\label{Tab:04}} {\begin{tabular}{ll}\hline
9-mer & Gene Name   \\ \hline
RAWRVVFEA & AEGP \\
APGPRGFPG & COAA1 \\
PTYTVWYLG  & FA43B \\
PSAVAHVVV  & FAT1 \\
LKEGEEDGR & MTA1 \\
PSKLHNYST & NMDE2 \\
QLAQLSSPC  & ROBO4 \\
MWALCSLLR  & ELAC2 \\
QAPGSVLFQ & SALL4 \\
MAGIRVTKV & TGM6 \\
PPVASFDYY & TNR \\
SLMRQKFQW & MRC1 \\
HVSNGAPVP & MSH2 \\
KRGYFDFRI  & PGBD1 \\
LNRGELKLI  & PIM1 \\
\hline
\end{tabular}}
\end{table}

\subsection{UniProtKB Human Gene binding prediction}

\begin{figure*}[tbh]
	\centering     
	\subfigure[human genome peptides]{\label{fig:a}\includegraphics[width=0.48\textwidth]{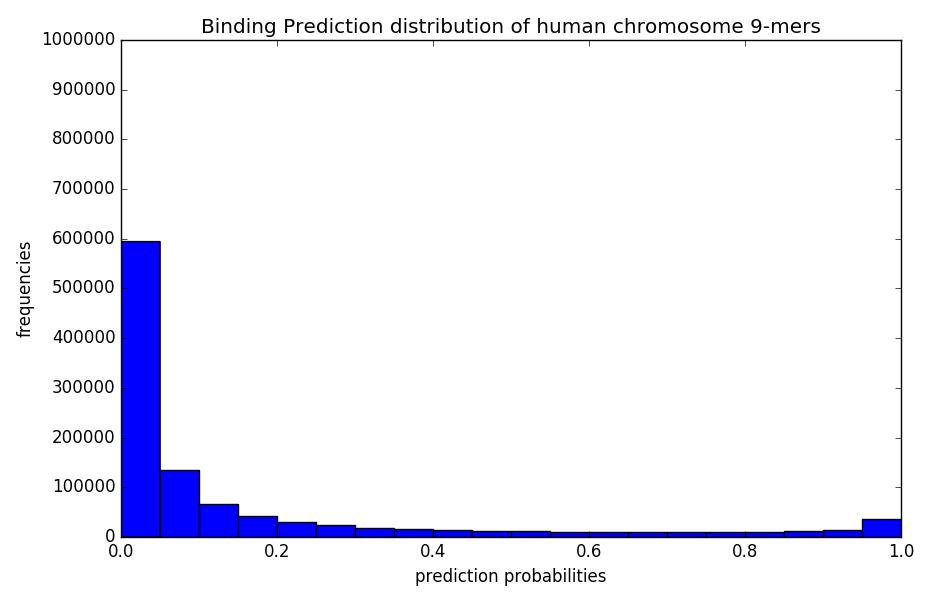}}
	\subfigure[artificially generated peptides]{\label{fig:b}\includegraphics[width=0.48\textwidth]{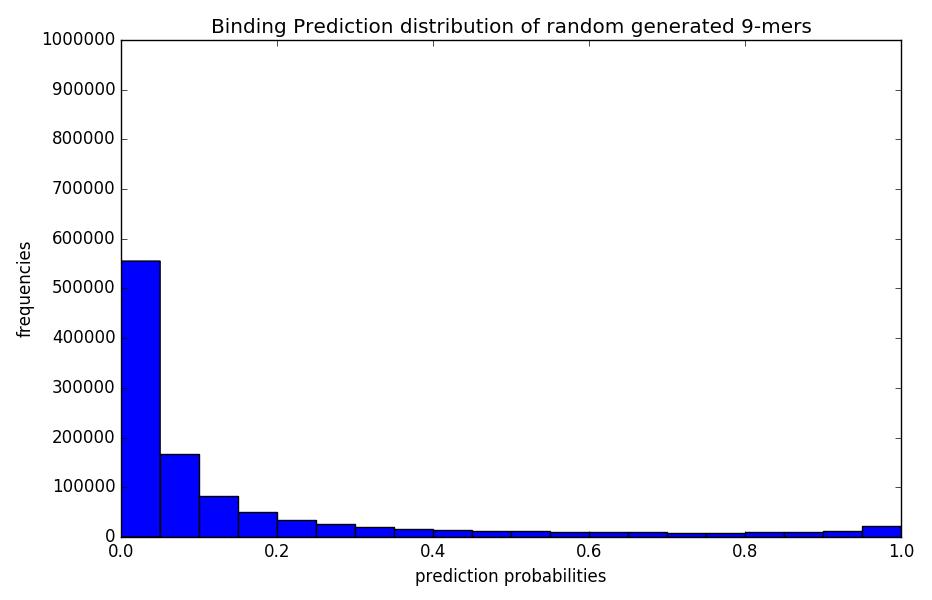}}
	\caption{Distributions of prediction binding probabilities using HLA-CNN trained on A*02:01 allele subtype. (a) shows the predicted distribution of human chromosome 9-mers. (b) shows the predicted distribution of random generated 9-mers.}\label{fig:04}
\end{figure*}

We perform binding prediction experiment on the entire 20,162 human protein-coding genome from UniProtKB \cite{UniProt} and randomly generated 9-mers.  The 20,162 human genes are chopped into 9-mers, with duplicates and those containing amino acids X, B, and U filtered out, leaving 1,0873,314 unique self 9-mers. An equivalent number of nonself 9-mer proteins, exclusive the self 9-mers, were obtained by randomly permuting the 20 amino acids.

The HLA class I - A gene alone is reported to have almost 4000 different alleles \cite{Marsh10}, each estimated to bind between 1,000 and 10,000 individual peptide sequences \cite{Brusic04}.  As each allele subtype is highly specific and binds to only a small subset of peptides that exhibits a particular motif \cite{Eisen12}, we were interested to see if any pattern could be discerned using our model to make binding predictions on the sets of self and nonself 9-mers. 

Fig. \ref{fig:04} shows the distributions of binding prediction for self and nonself 9-mers using HLA-CNN trained on the A*02:01 allele subtype. The distribution of predicted binding probablities between the two sets of self and nonself 9-mers are nearly identical. This was not unexpected as the small number of training data points compared to the overall size of the test sets led us to believe the model would exhibit similar level of false positives between the two 9-mer sets.

What is interesting is the fact that our model predicts a high number of potential self binding 9-mers.  Table \ref{Tab:04} shows the top 15 human 9-mers with highest predicted binding probabilities.  Shown next to each of these 9-mers are the name of the genes where these 9-mers originated from.  A literature review shows that these 15 genes are novel and not involved in any pathway of known auto-immune diseases. Our model indicates that these genes have the potential for self binding and may be worth validating in future experiments that are beyond the scope of this work.

\section{Conclusion}

In this work, we have described how machine learning techniques from the NLP domain could be applied to bioinformatics setting, specifically HLA class I-peptide binding prediction. We presented a method to extract a vector space distributed representation of amino acids from available HLA classI-peptide data that preserved property critical for covalent bonding. Using this vector space representation, we proposed a deep CNN architecture for the purpose of HLA class I-peptide binding prediction. This framework is capable of making prediction for any length peptides or any allele subtype, provided sufficient training data is available. Experimental results on the IEDB benchmark datasets demonstrate our algorithm achieved state-of-the-art binding prediction performance on the majority of test sets over existing models.

On future work, allele-specific affinity thresholds instead of a general binding affinity ic50 threshold of 500 nM can be used to identify peptide binders in different subtypes.  This approach had shown superior predictive efficacy in previous work \cite{Paul13}.  From an architecture design standpoint, one possibility to extend the network is to replace the dense layer with a convolutional layer, thereby creating a fully convolutional network (FCN).  The motivation being since convolutional layers preserve spatial information in the peptide, perhaps a FCN could improve performance over the existing network if all layers in the network had this capability. Another option is to generalize the single output architecture to multi-outputs.  Specifically, a secondary output layer and loss function can be added to minimize the mean square error between gold standard ic50 values and predicted ic50 values alongside the existing binary cross-entropy output layer. The underlying convolutional and fully connected layers would be shared between these two output layers/loss functions as the motivation would be to learn a model that has both good AUC quality as well as SRCC quality.

\end{document}